\documentclass[12pt]{article}
\usepackage{a4}    
\usepackage{amsmath,amssymb}

\begin{document}
\title{\bf Factorization of Unitary Matrices}
\author{P. Di\c t\u a\\
Institute of Physics and Nuclear Engineering,\\
P.O. Box MG6, Bucharest, Romania}
\date{}
\maketitle
\footnotetext{\noindent Electronic mail: dita@hera.theory.nipne.ro}

\begin{abstract}

Factorization of an $n\times n$ unitary matrix as a product of $n$ diagonal
matrices containing only phases interlaced with $n-1$ orthogonal matrices each
one generated by a real vector as well as an explicit form for the Weyl
factorization are found.  \end{abstract}


\newpage
\section{Introduction}
\label{intro}
Matrix factorization is a live subject of linear algebra. It seems that no
general theory is yet available although many results appear  almost every
day. However our goal will not be so ambitious to present a general theory of
matrix factorizations but to tackle the problem of factorization of unitary
matrices. Unitary matrices are a first hand  tool in solving many problems in
mathematical and theoretical physics and the diversity of the problems 
necessitates to keep improving it.
In fact the matrix factorization is closely related to the parametrization of
unitary matrices and the classical result by Murnagham on parametrization of
the $n$-dimensional unitary group $U(n)$ is the following: an arbitrary
$n\times n$ unitary matrix is the  product of a diagonal matrix containing $n$
phases and of $n(n-1)/2$ matrices  whose main building block has the form 
$$U=\left(\begin{array}{cc}
cos\,\theta&-sin\,\theta\,\,e^{-i\varphi}\\
sin\,\theta\,\,e^{i\varphi}&cos\,\theta\\
\end{array}\right)\eqno(1)$$
The parameters entering the parametrization are $n(n-1)/2$ angles $\theta_i$
and $n(n+1)/2$ phases $\varphi_i$. For more details see \cite{Mu}.

A selection of a specific set of angles and/or phases has no theoretical
significance because all the choices are mathematically equivalent; however a
clever choice may shed some light on important qualitative issues. This is the
point of view of Harari and Leurer \cite{HL} who recommended a standard choice
of the Cabbibo angles and Kobayashi-Maskawa phases for an arbitrary number of
quark generations and accordingly they propose a new parametrization
(factorization) which, up to a columns permutation, is  nothing else than
Murnagham parametrization with the change $\theta\,\rightarrow \,
\pi/2-\theta$ and $\varphi\rightarrow\varphi+\pi$.

However there are some problems that require a more elaborate factorization. 
To our knowledge one of the first such problems is that   raised by
Reck $et\,\,al. $ \cite{RZBB} who describe an experimental realization of any
discrete unitary operator. Such devices will find practical applications in
quantum cryptography and in quantum teleportation. Starting from  Murnagham
parametrization they show that  any $n\times n$ unitary matrix $A_n$ can be
written as a product
$$A_n=B_n\, C_{n-1}$$
where $B_n\, \in U(n)$ is at its turn a product of $n-1$ 
unitary matrices containing each one a block of the form (1)
and $C_{n-1}$ is a $U(n-1)$ matrix. Consequently the experimental realization
of a $n\times n$ unitary operator is reduced to the realization of
two unitary operators out of which one has a lesser dimension. The experimental
realization of a $U(3)$ matrix, sketched in Fig. 2 of their paper, suggests
that it would be preferable that  phases entering the parametrization
should be factored out, the device becoming simpler and the phase shifters, 
their terminology for phases, being placed at the input and output ports
respectively. 

A mixing of the Murnagham factorization and that of Reck $et \,\,al. $ is
that proposed by Rowe $et \,\,al. $ \cite{RSG}   in their study on  the
representations of Weyl group
and of Wigner functions for $SU(3)$.
The last parametrization is also used  by Nemoto in his attempt to develop
generalized coherent states for $SU(n)$ systems \cite{Ne}.

Another kind of factorization is that suggested by Chaturvedi and Mukunda in
their paper  \cite{CM} aiming at obtaining a more "suitable" parametrization
of the Kobayashi-Maskawa matrix. Although the proposed forms for $n=3,4$
are  awfully complicated by comparison with other parametrizations existing in
literature, and for this reason it cannot be extended easily to cases $n \ge 5
$, the paper 
contains a novel idea namely that that an $SU(n)$  matrix can be parametrized
by a sequence of $n-1$ complex    vectors of dimensions $2,3,\dots,n$.
Fortunately there is an alternative construction   as it may be inferred from
the construction of an $SU(3)$ matrix as a product of two matrices each of
them generated by a three- and respectively two-dimensional complex vector
\cite{MS} in a much more simple form than that presented in ref.\cite{CM}.

One aim of this paper is to  elaborate this alternative construction in
order to obtain a parametrization of $n\times n$ unitary matrices
as a product of $n$ diagonal matrices containing the phases and $n-1$
orthogonal matrices, each of them generated by a $n$-dimensional  real vector.
As a byproduct we obtain the Weyl form \cite{We} of a unitary matrix
$W=w^*\, d\,w$ where $w$ is a unitary matrix, $w^*$ its adjoint  and $d$ a
diagonal matrix containing phases. The Weyl factorization was the key
ingredient in finding the "radial" part of the Laplace-Beltrami operator on
$U(n)$ and $SU(n)$ \cite{Wa, MO} and this explicit form could help in finding
completely the Laplace-Beltrami operator on unitary groups.

The paper is organized as follows: In Sect. 2 we derive a factorization of
$n\times n$ unitary  matrices as a product of $n$ diagonal matrices
interlaced with $n-1$ orthogonal matrices generated by real vectors of
dimension $2,3\dots,n-1$.
The explicit form of orthogonal matrices is found in Sect.3 and the paper ends
with Concluding remarks.

\section{Factorization of unitary matrices}  
    
                             The unitary group $U(n)$ is the
group of automorphisms of the Hilbert space $({\bf C}^{n}, (\cdot,\cdot))$
where $(\cdot,\cdot)$ is the Hermitian scalar  product
$(x,y)=\sum_{i=1}^{i=n}\,\overline{x_i}\,y_i$. If $A_n\in  U(n)$ by $A_n^*$ we
will denote the adjoint matrix and then $A_n^*\,A_n=I_n$, where $I_n$  is
the $n\times n$ unit matrix. It  follows that $det\, A_n=
e^{i\,\varphi}$, where $\varphi$ is a phase, and $dim_{R}\,U(n)=n^2$.

First of all we want to introduce some notations that will be useful in the
following. The product of two unitary matrices being again a unitary matrix it
follows that the multiplication of a row or a column by an arbitrary phase
does not affect the unitarity property. Indeed the multiplication of the
$j^{th}$ row by $e^{i\,\varphi_j}$ is equivalent to the left multiplication by
a diagonal matrix whose all diagonal entries but the $j^{th}$ one are
equal to unity and $a_{jj}=e^{i\,\varphi_j}$. 
The first building blocks appearing in factorization  of unitary matrices
are diagonal matrices written in  the  form
$d_n=(e^{i\varphi_1},\dots, e^{i\varphi_n})$ with $\varphi_j \in
[0,2\,\pi),\,\,j=1,\dots,n$ arbitrary  phases, and all  off-diagonal
entries  zero.  We introduce also the notation
$d_k^{n-k}=(1_{n-k},e^{i\psi_1},\dots,e^{i\psi_k})$,  $k<n$, where
$1_{n-k}$ means that the first $(n-k)$ diagonal entries are equal to unity,
i.e. it can be obtained  from $d_n$ by making the first $n-k$ phases zero .
Multiplying at left by $d_n$ an arbitrary unitary matrix   the first  row
 will be multiplied by $e^{i\varphi_1}$, the second by
$e^{i\varphi_2}$, etc. and the last one by $e^{i\varphi_n}$. Multiplying
at right with $d_k^{n-k}$ the first $n-k$ columns remain unmodified and the
other ones are multiplied by $e^{i\psi_1}, \dots,e^{i\psi_k} $
respectively.  A consequence of this property is the following: the phases of
the elements of an arbitrary row and/or column can be taken zero or $\pi$ and
a convenient choice is to take the elements of first column  non-negative
numbers less than unity and those of the first row real numbers. This follows 
from the equivalence between the permutation of the $i^{th}$ and $j^{th}$ rows
(columns) with the left (right) multiplication by the unitary matrix $P_{ij}$
whose all diagonal entries but $a_{ii}$ and $a_{jj}$ are equal to unity,
$a_{ii}=a_{jj}=0,\, a_{ij}=a_{ji}=1,\, i\neq j$ and all the other entries
vanish. In conclusion an arbitrary $A_n\in  U(n)$  can be written as a product
of two matrices, the first one diagonal, in the form
$$A_n=d_n\,\tilde{A_n}\eqno(2)$$ where $\tilde{A_n}$ is a matrix with   
  the first  column entries non-negative numbers.

Other building blocks that will appear in factorization of $\tilde{A_n}$ 
are the rotations which operate in the $i,i+1$ plane of the form 
 $$J_{i,i+1}=\left(
\begin{array}{ccc}

I_{i-1}&0&0\\
0&
\begin{array}{cc}
cos\,{\theta_i}&-sin\,{\theta_i}\\
sin\,{\theta_i}& cos\,{\theta_i}
\end{array}
&0\\
0&0&I_{n-i-1}
\end{array}\right),\quad i=1,\dots,n-1\eqno(3)$$ 
 The above formula contains the block (1) with  phase zero unlike other
parametrizations \cite{Mu}-\cite{Ne}, in our parametrization the phases will
appear only in diagonal matrices.

 Let $v$ be the vector $v=(1,0,\dots,0)^t\,\in
S_{2n-1}\,\in{\bf C}^{n}$ where $t$ denotes transpose and $S_{2n-1}$ is       
                                           the unit sphere of the Hilbert space
${\bf C}^{n}$ whose  real dimension is $2n-1$. By applying  $A_n\in 
U(n)$ to the vector $v$ we find

$$ A_n\,v=a=\left( \begin{array}{c}
a_{11}\\
\cdot\\
\cdot\\
\cdot\\
a_{n1}
\end{array} \right)$$
where $a\in S_{2n-1}$  because $A_n$ is unitary. The vector $a$ is
completely determined by the first column of the matrix $A_n$.  Conversely,
given an arbitrary vector of the unit sphere $w\in S_{2n-1}$ this point
determines a unique first row of a unitary matrix  which maps $w$
to the vector $v$. Therefore $U(n)$ acts transitively on  $S_{2n-1}$.
The subgroup of $U(n)$ which leaves $v$  invariant is $U(n-1)$ on the last
$n-1$ dimensions  such that $$ S_{2n-1}={\it coset~space }\,\,U(n)/U(n-1)$$ 

A
direct consequence of the last relation is that we expect that any element of
$U(n)$ should be uniquely specified by a pair of a vector $b\in S_{2n-1}$ and
of an arbitrary element of $U(n-1)$. Thus we are looking for a factorization
of an arbitrary element $A_n\in U(n)$ in the form
$$A_n=B_n\cdot\left(\begin{array}{cc}
1&0\\0&A_{n-1}\end{array}\right)\eqno(4)$$
 where $B_n\in U(n)$ is a unitary
matrix whose first column is uniquely defined by a vector $b\in S_{2n-1}$, but
otherwise still arbitrary and $A_{n-1}$ is an arbitrary element of $U(n-1)$.
For the $SU(3)$ group such a factorization was obtained recently \cite{CM,MS}.
Iterating the previous equation we arrive at the conclusion that an element of
$U(n)$ can be written as a product of $n$ unitary matrices 
                             
$$A_n=B_{n}\cdot B_{n-1}^1\dots B_1^{n-1}\eqno(5)$$
where
$$B_{n-k}^k=\left(\begin{array}{cc}
I_k&0\\
0&B_{n-k} \end{array}\right)$$
$B_k,\,\, k=1,\dots,n-1$ ,are $k\times k$ unitary matrices whose
first column is generated by vectors $b_k\in S_{2k-1}$; for
example $B_1^{n-1}$ is the diagonal matrix
$(1,\dots,1,e^{i\varphi_{n(n+1)}})$.

The still arbitrary columns of $B_k$ will be chosen in such a way that we
should obtain a simple form for the matrices $B^{n-k}_k$, and we 
require that  $B_k$ should be completely specified by the parameters entering
the vector $b_k$ and nothing else. In the following we show that such a
parametrization  does exist  and then   $A_n\in U_n$ in Eq.(5) will be
written as a product of $n\times n$ unitary  matrices each one
parametrized by $2k-1,\,\, k=1,\dots,n$, real parameters such that the number
of independent parameters entering $A_n$ will be $1+3+\dots+2n-1=n^2$ as  it
should  be.

In other words our problem is to complete an $n\times n$  matrix
whose its first column is given by a vector $b_n\in S_{2n-1}$ to a unitary
matrix and we have to do it without introducing supplementary parameters. For
$n=3$ this was found by us in \cite{Di} in an other context and here
we give  the construction for arbitrary $n$.

If we take into account the property (2) the problem simplifies a little 
since then
$$B_n=d_n\,\tilde{B}_n$$
where the first column of $\tilde{B}_n$ has non-negative entries. Denoting
this column by $b_1$ we will use the parametrization 
$$b_1=(cos\,{\theta_1},cos\,{\theta_2}\,sin\,{\theta_1
},\dots,\sin\,{\theta_1}\dots sin\,{\theta_{n-1}})^t \eqno(6)$$
where $\theta_i\in[0,\pi/2],\, i=1,\dots,n-1$; we call $\theta_i$ angles. Thus
$B_n$ will be parametrized by $n$ phases and $n-1$ angles. According to the
above factorization $\tilde{B}_n$ is nothing else than the orthogonal matrix
generated by the vector $b_1$. Thus with no loss of generality
$B_n=d_n\,{\cal{O}}_n$ with ${\cal{O}}_n\in O(n)$.
In this way the factorization of $A_n$ will be
$$A_n=d_n\,{\cal{O}}_n\,d_{n-1}^1\,{\cal{O}}_{n-1}^1\dots
d_2^{n-2}{\cal{O}}_{2}^{n-1}d_1^{n-1}I_n$$
where ${\cal{O}}_{n-k}^k$ has the same structure as $B_{n-k}^k$, i.e
$$ {\cal{O}}_{n-k}^k=\left(\begin{array}{cc}
I_k&0\\
0&{\cal{O}}_{n-k}
\end{array}\right)$$
Consequently the factorization of unitary matrices reduces to the
parametrization of  orthogonal matrices generated by an arbitrary vector of
the real $n$-dimensional sphere and in the next section we show how to
do it.
\section{Parametrization of orthogonal matrices}
An operator $T$ applying the Hilbert space $\cal{H}$ in the Hilbert space
$\cal{H}'$ is a contraction if for any $v\in
{\cal{H}}$,~~$||T\,v||_{{\cal{H'}}}\leq ||v||_{{\cal{H}}}$, i.e. $||T||\leq
1$, \cite{FN}. For any contraction we have $T^*\,T\leq I_{{\cal{H'}}}$ and
$T\,T^*\leq I_{{\cal{H}}}$ and the defect operators
$$D_T=(I_{{\cal{H}}}-T^*\,T)^{1/2},\quad
D_{T^*}=(I_{{\cal{H'}}}-T\,T^*)^{1/2}$$ are Hermitian operators  in
${{\cal{H}}}$ and ${{\cal{H}'}}$ respectively. They have the property
$$ T\,D_{T}=D_{T^*}\,T,\quad T^*\,D_{T^*}=D_{T}\,T^*\eqno(7)$$

In the following we are interested in a contraction of a special form,
namely that generated by a $n$-dimensional real vector $b\in {\bf R^n}$, i.e.
$T=(b_1,\dots,b_n)^t$, where $b_i$ are the coordinates of $b$; its norm is
$||T||=(b,b)$ and $T$ will be a contraction iff $(b,b)\leq 1$, i.e. if $b$  is
a point inside the unit ball of ${\bf R}^n$. If $(b,b)= 1$, that is the case
we are interested in, $T$ is an isometry, i.e.
$$T^*\,T=1,\qquad{\rm and}~~D_T=0$$
and in this case $D_{T*}$ is an orthogonal projection. A direct calculation
shows that $det\,(\lambda I_n-D_{T^*}^2)=\lambda(\lambda -1)^{n-1}$ such that
the eigenvalue $\lambda=0$ has unit multiplicity and the eigenvalue
$\lambda=1$ is degenerated. From the second relation (7) we have
$$D_{T^*}\,T=D_{T^*}\,b=T\,D_T=b\,D_T=0$$
i.e. $b$ is the eigenvector of $ D_{T^*}$ corresponding to $\lambda=0$
eigenvalue.

The orthogonal matrix ${\cal{O}}_n$ which brings the operator $D_{T^*}$ 
to a diagonal form 
$${\cal{O}}_n^t\, D_{T^*}{\cal{O}}_n=\left( \begin{array}{cc}
0&0\\
0&I_{n-1}
\end{array} \right)$$
is the orthogonal matrix we are looking for  because it is generated by an
arbitrary $n$-dimensional real vector of unit norm. The multiplicity of
$\lambda=1$ eigenvalue being $n-1$ the form of the matrix ${\cal{O}}_n$ is not
uniquely defined. Indeed if $v_1,\dots,v_n$ are the eigenvectors of $D_{T^*}$ 
$$D_{T^*}v_1=D_{T^*}b=0,\quad D_{T^*}v_k=v_k,\quad k=2,\dots,n\eqno(8)$$
orthogonal eigenvectors will be also the vectors 
$$v_1 ~~{\rm and}~~a_k=R\,v_k,\quad k=2,3,\dots,n$$
where $R \in O(n)$ is an arbitrary 
rotation acting only on the last $n-1$ eigenvectors. Thus there is a continuum
of solutions for the orthogonal basis that diagonalizes $D_{T^*}$. In this
situation we have to make a choice between the possible bases. Our  criteria
was that the resulting orthogonal matrix ${\cal{O}}_n$ should have as many as
possible vanishing entries. We found such a  matrix that have $(n-1)(n-2)/2$
null entries and the result is expressed by the following lemma:

{\bf Lemma 1}:{\it The eigenvectors of the eigenvalue problem Eq.(8) which
are the columns of the orthogonal matrix ${\cal O}_n \in SO(n)$ generated by
the vector parametrized by Eq.(6) are given by

$$v_1=\left(\begin{array}{l}
cos\,\theta_1\\
sin\,\theta_1\,cos\,\theta_2\\
\cdot\\
\cdot\\
\cdot\\
sin\,\theta_1\,\dots sin\,\theta_{n-1}
\end{array} \right),\,\,
v_{2}=\left(\begin{array}{l}
-sin\,\theta_1\\
cos\,\theta_1\,cos\,\theta_2\\
cos\,\theta_1\,sin\,\theta_2\,cos\,\theta_3\\
\cdot\\
\cdot\\
cos\,\theta_1\,sin\,\theta_2\dots sin\,\theta_{n-1}
\end{array}\right)$$
$$\eqno(9)$$
$$v_{k+2}=\left(\begin{array}{l}
0_k\\
-sin\,\theta_{k+1}\\
cos\,\theta_{k+1}\,cos\,\theta_{k+2}\\
\cdot\\
\cdot\\
cos\,\theta_{k+1}\,sin\,\theta_{k+2}\dots
sin\,\theta_{n-1}\end{array}\right),\qquad k=1,\dots,n-2
$$
where $0_k$ means that all the first $k$ entries are zero and $k=1,2,\dots
n-2$. Alternatively 
$$v_{k+1}={d\over d\,\theta_k }\,\,v_1(\theta_1=\dots
=\theta_{k-1}=\pi/2)\qquad k=1,\dots,n-1$$
The full $O(n)$ group is obtained by multiplying ${\cal O}_n$ given by Eqs.(9)
with the diagonal matrix which has all the entries but  the last one $1$ and
$d_{n,n}=-1$.} 

{\it
Proof}: Elementary checking shows that $(v_i,v_j)=\delta_{ij},\,\,
i,j=1,\dots,n$, and thus $v_k$ are linearly independent. Because the
multiplicity of the null eigenvalue is unity it follows that $v_k,\,\,
k=2,\dots,n$, are orthogonal eigenvectors corresponding to $\lambda=1$
eigenvalue.

{\bf Lemma 2} {\it The orthogonal matrices ${\cal O}_n$ ( ${\cal O}_{n-k}^k$)
at their turn can be  factored into a product of $n-1$ (n-k-1)
matrices of the form $J_{i,i+1}$; e.g. we have 
$${\cal O}_n=J_{n-1,n}\,J_{n-2,n-1}\dots
J_{1,2}\eqno(10)$$
where $J_{i,i+1}$ are $n\times n$ rotations introduced by Eq.(6). }

{\bf Remark}. If the angles that parametrize ${\cal O}_n$ are
$\theta_1,\dots,\theta_{n-1}$, then the angles that parametrize ${\cal
O}_{n-1}^1$ are denoted e.g. by $\theta_n,\dots,\theta_{2n-3}$, etc. and the
last angle entering ${\cal
O}^{n-1}_2$ will be $\theta_{n(n-1)/2}$.

 Putting together all the preceding
information one obtains  the following
result

{\bf Theorem}: {\it Any  element $A_n\in U(n)$ can be factored into an ordered
product of $2n-1$ matrices of the following form
$$A_n=d_n\,{\cal{O}}_n\,d_{n-1}^1\,{\cal{O}}_{n-1}^1\dots
d_2^{n-2}{\cal{O}}_{2}^{n-1}d_1^{n-1}\eqno(10)$$
 where $d_{n-k}^k$ are diagonal
matrices and  ${\cal{O}}_{n-k}^k$ orthogonal matrices whose columns are given
by formulae like (9)  generated by real $(n-k)$-dimensional unit vectors.
Using  factorization (10) the above formula can be written as a product of
$n$ diagonal matrices and of $n(n-1)/2$ rotations $J_{k,k+1}$.

The condition $\sum_{i=1}^{n(n+1)/2}\,\varphi_i=0$, imposed on  $\varphi_i $
 the arbitrary phases entering the parametrization of $A_n$, gives  the
factorization of $SU(n)$ matrices.

   If
$w_n={\cal{O}}_n\,d_{n-1}^1\,{\cal{O}}_{n-1}^1\dots
d_2^{n-2}{\cal{O}}_{2}^{n-1}d_1^{n-1}={\cal{O}}_n\,d_{n-1}^1\,w_{n-1}$ then
$$W_n=w_n^*\,d_nw_n\eqno(11)$$ is one (of the many possible) Weyl
representation of unitary matrices. 

If all the phases   entering $A_n$ are zero $\varphi_i=0,\, i=1,\dots,
n(n+1)/2$, one gets the factorization of rotations $R_n \in
SO(n)$ 
$$R_n={\cal{O}}_n\,{\cal{O}}_{n-1}^1\dots{\cal{O}}_{2}^{n-1}\eqno(12)$$
and the full group $O(n)$ is obtained by multiplying (12) with a diagonal matrix
$d=(1,\dots,1,-1)$ that has one entry equal to $-1$.}

{\it Remark}. The above factorization is not unique and we propose it as the
standard (and simplest) representation. Equivalent
factorizations (parametrizations) can be obtained by inserting matrices like 
$P_{ij}$ as factors in the formulae (10)-(12) since the number of parameters
remains the same and only the final form of the matrices will be different.
As concerns Eq.(11) we made the choice that leads to the simplest form for the
matrix elements of $W_n$ as polynomial functions of sines and cosines which
enter the parametrization of orthogonal matrices. For example instead of 
$w_n={\cal{O}}_n\,d_{n-1}^1\,w_{n-1}$ we could take
$w_n={\cal{O}}_n\,W_{n-1}$, where $W_{n-1}$ is at its turn given by
a formula like Eq.(11) and so on.

{\it Examples}. An element $A_4 \in U(4)$ factors as
$$A_4=d_4\,{\cal{O}}_4\,d_{3}^1\,{\cal{O}}_{3}^1\,d_{2}^2\,{\cal{O}}_{2}^2\,d_{1}^3$$
where $d_4=(e^{i \varphi_1},e^{i \varphi_2},e^{i \varphi_3},e^{i
\varphi_4})\,\, d_3^1=(1,e^{i \varphi_5},e^{i \varphi_6},e^{i \varphi_7})\,\, 
d_2^2=(1,1,e^{i \varphi_8},e^{i \varphi_9})$, 
$d_1^3=(1,1,1,e^{i \varphi_{10}})$ and ${\cal{O}}_4$, ${\cal{O}}_3^1$ and
${\cal{O}}_2^2$ are the following matrices 

$${\cal O}_4=\left(\begin{array}{llcc}
cos\,{\theta_1}& -sin\,{\theta_1}&0&0\\
sin\,\theta_1\,cos\,\theta_2&cos\,\theta_1\,cos\,\theta_2&
-sin\,\theta_2&0\\
sin\,\theta_1\,sin\,\theta_2\,cos\,\theta_3&cos\,\theta_1\,
sin\,\theta_2\,cos\,\theta_3&cos\,\theta_2\,cos\,\theta_3&-sin\,\theta_3\\
sin\,\theta_1\,sin\,\theta_2\,sin\,\theta_3&cos\,\theta_1\,sin\,\theta_2\,
sin\,\theta_3&cos\,\theta_2\,sin\,\theta_3&cos\,\theta_3\\
\end{array}\right)$$

$${\cal{O}}_3^1=\left(\begin{array}{lccc}
1&0&0&0\\
0&cos\,\theta_4& -sin\,\theta_4&0\\
0&sin\,\theta_4\,cos\,\theta_5&cos\,\theta_4\,cos\,\theta_5&-sin\,\theta_5\\
0&sin\,\theta_4\,sin\,\theta_5&cos\,\theta_4\,sin\,\theta_5&cos\,\theta_5\\
\end{array}\right)$$

$${\cal{O}}_2^2=\left(\begin{array}{lccc}
1&0&0&0\\
0&1&0&0\\
0&0&cos\,\theta_6& -sin\,\theta_6\\
0&0&sin\,\theta_6&cos\, \theta_6\\
\end{array}\right)$$

The formula (10) for ${\cal{O}}_4$ is

$${\cal{O}}_4=J_{3,4}\cdot J_{2,3}\cdot J_{1,2}=$$
$$
\left(\begin{array}{cccc}
1&0&0&0\\
0&1&0&0\\
0&0&cos\,\theta_3&-sin\,\theta_3\\
0&0&sin\,\theta_3&cos\,\theta_3\\
\end{array}\right)
\left(\begin{array}{cccc}
1&0&0&0\\
0&cos\,\theta_2&-sin\,\theta_2&0\\
0&sin\,\theta_2&cos\,\theta_2&0\\
0&0&0&1\\ \end{array}\right)
\left(\begin{array}{cccc}
cos\,\theta_1&-sin\,\theta_1&0&0\\
sin\,\theta_1&cos\,\theta_1&0&0\\
0&0&1&0\\
0&0&0&1\\
\end{array}\right)
$$
The Weyl form of a $2\times 2$ unitary  matrix is
$$W_2=w^*_2\,d_2\,w_2={d_1^1}^*\,{\cal O}_2^t\,d_2\,{\cal O}_2\,d_1^1=$$
$$\left(\begin{array}{cc}
e^{i\varphi_1}\,\,cos^2\, \theta+e^{i\varphi_2}\,\,sin^2\, \theta&
cos\, \theta\,\, sin\, \theta
\,\,e^{i\varphi_3}(e^{i\varphi_1}-e^{i\varphi_2})\\
&\\
 cos\,
\theta\,\, sin\, \theta \,\,
e^{-i\varphi_3}(e^{i\varphi_1}-e^{i\varphi_2})&e^{i\varphi_2}\,\,cos^2\,
\theta+e^{i\varphi_1}\,\,sin^2 \theta\\
\end{array}\right)$$
where $d_2=(e^{i\varphi_1},e^{i\varphi_2})$, $d_1^1=(1,e^{i\varphi_3})$ and $
{\cal O}_2=U$, and $U$ is the matrix (1).
\section{Concluding remarks}
In this paper we proposed a new factorization of unitary matrices which can be
useful in many domains of mathematical and theoretical physics. For example
the construction of coherent states for $SU(n)$ can be developed similarly to
that for the $SU(3)$ group given in \cite{MS}. The $SU(n)$ matrices are
obtained from the above parametrization by imposing the condition
$\sum_{i=1}^{i=n(n-1)}\,\varphi_i=0$ upon the phases and let us denote by the
same letter $A_n$ an arbitrary element of $SU(n)$. We consider the first $n-1$
columns of $A_n$  as vectors $v_k$ whose entries are given by 
$$(v_k)_i=A_{i,k}\,,\qquad i=1,\dots,n,\,\,k=1,\dots,n-1$$
and consider a set of $n(n-1)$ annihilation operators that we view as the
components of   $(n-1)$ annihilation vector operators
$$(a_k)_i=a_{ik},\qquad i=1,\dots,n,\,\,k=1,\dots,n-1$$ with the usual
commutation relations
$$[a_{ik}\,,a_{jl}]=0,\quad
[a_{ik}\,,a_{jl}^{\dagger}]=\delta_{ij}\,\delta_{kl}$$ The key element in
defining coherent states is the generating function
$$|v_1,\dots,v_{n-1}>_{n_1,\dots,n_{n-1}}=\sqrt{n_1\,!\dots\,n_{n-1}!}\,\,
exp\,\sum_1^{n-1}(a^{\dagger}_i,v_i)\,|0>$$ where $n_1,\dots,n_{n-1}$
are the integers that index the representation, $|0>$ is the vacuum vector and
we used the standard notation ${\dagger}$ for the adjoint of the annihilation
operator.

Another problem could be the finding of Laplace-Beltrami
operators on unitary groups that is an old problem { \cite{BR}}. The
Laplace-Beltrami operator on $S_{2n-1}$ can be written easily as

$$\Delta=\sum_{k=1}^{n-1}{1\over
sin^2\theta_1\dots sin^2\theta_{k-1}\,cos\,\theta_{k}sin^{2(n-k)-1}\theta_{k}}
{\partial\over\partial\theta_{k}}(cos\,\theta_k\,sin^{2(n-k)-1}\theta_k
{\partial\over\partial\theta_{k}} ) +$$
$$\sum_{k=1}^{n-1}{1\over
sin^2\theta_1\dots
sin^2\theta_{k-1}\,cos^2\theta_{k}}{\partial^2\over\partial\varphi^2_k}+{1\over
sin^2\theta_1\dots sin^2\theta_{n-1}}{\partial^2\over\partial\varphi^2_n}$$
where we used the parametrisation $v_n=({e^{\varphi_1}cos\,\theta_1,\dots,
e^{\varphi_n}}sin\,\theta_1\dots sin\,\theta_{n-1})\in S_{2n-1}$
With factorization (10) the mathematical tractability problem for $SU(3)$
\cite{BR} and other unitary groups can be resolved. 
A complete treatment of such problems will be given elsewhere.

Our proposal is largely
based on several simple ideas suggested earlier by many people and we have
written the factorization in the simplest possible way. We suggest it to
become the standard one since any other existing parametrization can be
brought to  this form  by multiplying      with permutation matrices like
$P_{ij}$ and/or diagonal matrices with entries containing phases.
\vskip1cm
\noindent
{\bf Acknowledgments}

The author acknowledges a partial financial support of the  Rumanian Academy
through the grant No.49/2000.

\end{document}